\documentclass[dvips,twoside,fleqn]{article}
\usepackage{times}
\usepackage{epsfig}
\usepackage{amstext}
\usepackage{amssymb}
\usepackage{espcrc2}

\title{Lattice Background Effective Action: a Proposal}

\author{
Paolo Cea\address{Dipartimento di Fisica - Universit\`a di Bari,
               via Amendola 173,
               70126 Bari, Italy}$^{\text{,b}}$
and Leonardo Cosmai\address{INFN - Sezione di Bari,
               via Amendola 173, 70126 Bari, Italy}
}

\begin{document}

\begin{abstract}
We propose a method based on the Schr\"odinger functional
for computing on the lattice the gauge invariant effective action for external
background fields. We check this method by studying 
the U(1) lattice gauge theory in presence of a constant magnetic background
field. 
\end{abstract}

\maketitle

\section{INTRODUCTION}

We propose a method, based on the Euclidean Schr\"odinger
functional~\cite{Symanzik81}, to evaluate on the lattice the gauge
invariant effective action. The Euclidean Schr\"odinger functional in
Yang-Mills theories without matter fields is:
\begin{equation}
\label{Eq1}
{\mathcal{Z}} \left[ A^{(f)}, A^{(i)} \right] =
 \langle  A^{(f)}  |  \exp(-HT) {\mathcal{P}} |
A^{(i)}
\rangle
\end{equation}
where ${\mathcal{P}}$ is the operator that projects onto the physical
states and $H$ the pure gauge Yang-Mills Hamiltonian in the temporal
gauge~\cite{Gross81}.

The Schr\"odinger functional on the lattice
becomes~\cite{Wolff86,Luscher92}:
\begin{equation}
\label{Eq2}
{\mathcal{Z}} \left[ U^{(f)}, U^{(i)} \right] =
\int {\mathcal{D}}U \exp(-S)
\,.
\end{equation}
with $S$ the Wilson action
modified to take into account the boundaries at $x_4=0$, and
$x_4=T$:
\begin{equation}
\label{Eq5}
U(x)|_{x_4=0} = U^{(i)} \, ,  \quad  U(x)|_{x_4=T} = U^{(f)} \,.
\end{equation}
The Schr\"odinger functional is 
invariant under arbitrary lattice gauge
transformations of the boundary links.

\section{LATTICE EFFECTIVE ACTION}

We want to investigate the lattice effective action for external
background fields.

In our approach~\cite{Cea96} we use periodic boundary conditions also in
the time direction:
\begin{equation}
\label{Eq6}
U(x)|_{x_4=0} = U(x)|_{x_4=T} = U^{\text{ext}}(0,\vec{x}) \,,
\end{equation}
so that $S$ is the standard Wilson action $S_W$.

On the lattice (${\text{P}}$ is the path-ordering operator):
\begin{equation}
\label{Eq7}
U_\mu^{\text{ext}}(x) = {\text{P}} \exp \left\{ + iag  \int_0^1 dt \,
A_\mu^{\text{ext}}(x+ at {\hat{\mu}}) \right\} \,,
\end{equation}
with the continuum gauge field ($\lambda_a/2$ generators of the SU(N)
Lie algebra):
\begin{equation}
\label{Eq8}
\vec{A}^{\text{ext}}(\vec{x}) =  \vec{A}_a^{\text{ext}}(\vec{x})
\lambda_a/2   \,.
\end{equation}
The lattice effective action for the background field
$A_\mu^{\text{ext}}(\vec{x})$ is defined by means of the  lattice
Schr\"odinger functional Eq.~(\ref{Eq2})
\begin{equation}
\label{Eq9}
\Gamma\left[ \vec{A}^{\text{ext}} \right] = -\frac{1}{T}
\ln \left\{ \frac{{\mathcal{Z}}[U^{\text{ext}}]}{{\mathcal{Z}}(0)}
\right\} \,,
\end{equation}
where $T$ is the extension in Euclidean time and
\begin{equation}
\label{ZUext}
{\mathcal{Z}}[U^{\text{ext}}]= 
{\mathcal{Z}}[U^{\text{ext}},U^{\text{ext}}]=
\int {\mathcal{D}} U \, \exp(-S_W)    \,.
\end{equation}
${\mathcal{Z}}(0)$ is the lattice Schr\"odinger functional
without external background field  (i.e. with $U_\mu^{\text{ext}}
=1$).
In the continuum, where $T\rightarrow\infty$, 
$\Gamma[\vec{A}^{\text{ext}}]$ becomes the
vacuum energy in presence of the background
field $\vec{A}^{\text{ext}}(\vec{x})$.

Our effective action is by definition gauge invariant
and can be used for a non-perturbative
investigation of the properties of the quantum vacuum.

\section{U(1) IN A CONSTANT BACKGROUND FIELD}

As a first step we check the consistency of our proposal by analyzing
the well known  U(1) lattice gauge theory.

We consider background fields that give rise to  constant field
strength. In this case $\Gamma[ \vec{A}^{\text{ext}}]$ is
proportional to the spatial volume  $V$ and the relevant quantity is 
the density of the effective action:
\begin{equation}
\label{Eq11}
\varepsilon\left[ \vec{A}^{\text{ext}} \right] =
-\frac{1}{\Omega} \ln \left[
\frac{{\mathcal{Z}}[U^{\text{ext}}]}{{\mathcal{Z}}(0)} \right] 
\,, \quad \Omega=V \cdot T \,.
\end{equation}

We study the U(1) l.g.t. in a constant background magnetic field 
directed along the  $x_3$ direction. 
In the Landau gauge: 
\begin{equation} 
\label{Eq12}
A_k^{\text{ext}}(\vec{x}) = \delta_{k,2} x_1 B  \,.
\end{equation} 
On the lattice: 
\begin{eqnarray}
\label{Eq13} 
\lefteqn{U_2^{\text{ext}}(x) = \exp \left[ i a g B x_1 \right]  \,,}
\nonumber \\
\lefteqn{U^{\text{ext}}_1(x) =  U^{\text{ext}}_3(x) =
U^{\text{ext}}_4(x) = 1  \,.}
\end{eqnarray} 
Since we adopt periodic boundary
conditions the external magnetic field gets quantized: 
\begin{equation} 
\label{Eq14} 
a^2 g B = \frac{2 \pi}{L_1} n^{\text{ext}} 
\end{equation} 
with $n^{\text{ext}}$ integer and
$L_1$ the lattice extension in the  $x_1$ direction (in lattice
units).

We perform numerical simulations of U(1) l.g.t. with the standard
Wilson action.  The links belonging to the time slice $x_4=0$ are
frozen to the configuration~(\ref{Eq13}). We also impose that the
constraint~(\ref{Eq13}) applies to links at the spatial boundaries
(in the continuum this condition amounts to the usual requirement that
the fluctuations over the background field vanish at the infinity).

We first analyze the behaviour of the  magnetic field.
To this end we look at the field strength tensor measured at a 
given time slice:
\begin{equation}
\label{field}
F_{\mu\nu}(x_4)= \sqrt{\beta} \, \left\langle \frac{1}{V}
\sum_{\vec{x}} \sin \theta_{\mu\nu} (\vec{x}, x_4) \right\rangle
\,.
\end{equation}
Only the component $F_{12}$ of the
field strength tensor is present in our data. Moreover,
in agreement with previous studies~\cite{DeGrand80}, 
we find that in the confined region $\beta <1$ the external
magnetic field is shielded after a small penetration,
while in the Coulomb region $\beta > 1$ the field penetrates 
indicating that the gauge system supports a long range magnetic field.

Let us turn now to the evaluation of the density of the effective action
Eq.~(\ref{Eq11}).
In this case we are faced with the problem of computing a partition  
function. 
To overcome this problem we consider the derivative of
$\varepsilon[\vec{A}^{\text{ext}}]$
with respect to $\beta$:
\begin{eqnarray}
\label{density}
\lefteqn{\varepsilon^{\prime} \left[ \vec{A}^{\text{ext}} \right]  = 
\frac{\partial \varepsilon \left[ \vec{A}^{\text{ext}}
\right]}{\partial \beta} = }
\\
\lefteqn{   
\left\langle \frac{1}{\Omega} \sum_{x,\mu>\nu} \cos \theta_{\mu\nu}(x)
\right\rangle_0 -
\left\langle \frac{1}{\Omega} \sum_{x,\mu>\nu} \cos \theta_{\mu\nu}(x)
\right\rangle_{A^{\text{ext}}} 
\,.}    \nonumber
\end{eqnarray}
The density of the effective action can be
recovered by integrating 
$\varepsilon^{\prime}$ over $\beta$.
Note that the contributions to  $\varepsilon^{\prime} [
\vec{A}^{\text{ext}} ]$  due to the frozen time slice at
$x_4=0$ and to the fixed boundary conditions at the lattice spatial
boundaries must be subtracted, i.e.
only the dynamical
links must be taken into account in evaluating
$\varepsilon^{\prime} [\vec{A}^{\text{ext}} ]$.
We denote by  $\Omega=L_1 L_2 L_3 L_4$ the total number of lattice
sites (i.e. the lattice volume).
$\Omega_{\text{ext}}$ are the lattice sites whose links
are fixed according to Eq.~(\ref{Eq6}):
\begin{eqnarray}
\label{Omegaext}
\lefteqn{\Omega_{\text{ext}}   =    L_1 L_2 L_3 +  (L_4-1)}  \nonumber  \\
\lefteqn{\,\, \times  (L_1 L_2 L_3 - (L_1-2)(L_2-2)(L_3-2)) \,.}
\end{eqnarray}
Hence $\Omega_{\text{int}}=\Omega-\Omega_{\text{ext}}$ is the
volume occupied by the  dynamical lattice sites.
In Figure~1 we display the derivative of the energy density due to
``internal'' links versus $\beta$ for the $64\times 12^3$ lattice and
$n^{\text{ext}}=2$. 
\begin{figure}[tbp]
\begin{center}
\epsfig{file=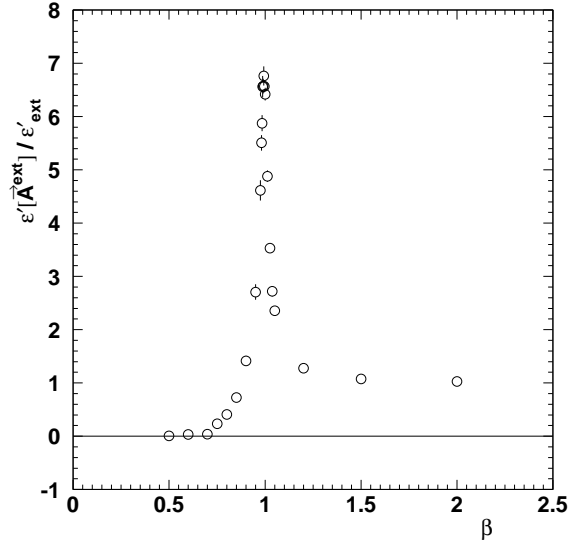,width=7.5truecm}
\vspace{-1.54truecm}
\caption{The derivative of the energy density due to links belonging
to $\Omega_{\text{int}}$ versus $\beta$ for the $64 \times 12^3$
lattice.}
\end{center}
\end{figure}
\begin{figure}[tbp]
\begin{center}
\epsfig{file=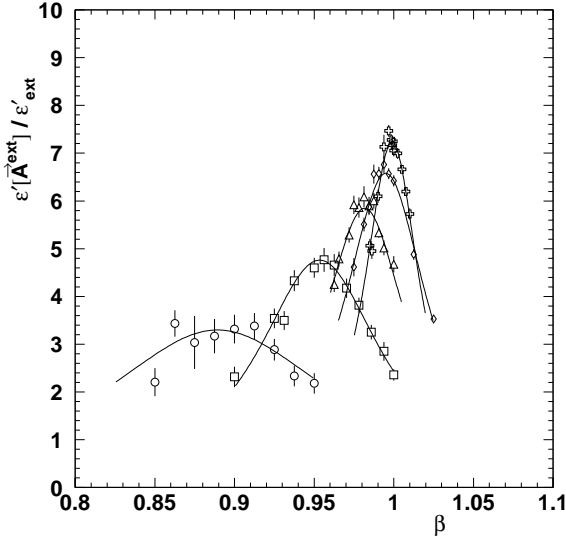,width=7.5truecm}
\vspace{-1.54truecm}
\caption{$\varepsilon_{\text{int}}^{\prime}$ near the critical region
for the lattices $64 \times 6^3$ (open circles),  $64 \times 8^3$
(open squares), $64 \times 10^3$ (open triangles),   $64 \times 12^3$
(open diamonds), and $64 \times 14^3$ (open crosses), with the
superimposed fits using Eq.~(\ref{peak}).}
\end{center}
\end{figure}

Figure~1 shows that 
in the weak coupling region 
($\beta \gg 1$)
$\varepsilon^{\prime}[ \vec{A}^{\text{ext}}]$
tends to the derivative of the external action (action due to the external
links):
\begin{equation}
\label{epspext}
\varepsilon^{\prime}_{\text{ext}} = \frac{\partial}{\partial \beta}
\frac{1}{\Omega} S^{\text{ext}} = 1- \cos \left( \frac{2 \pi}{L_1}
n^{\text{ext}} \right) \,.
\end{equation}
This means that for large $\beta$
the effective action agrees with the classical action:
\begin{eqnarray}
\label{classact}
\lim_{\beta \rightarrow \infty} \varepsilon [\vec{A}^{\text{ext}}]
& = &  \beta
\left[ 1 - \cos \left( \frac{2 \pi}{L_1} n^{\text{ext}} \right)
\right] 
\nonumber  \\
& = &
\varepsilon_{\text{ext}} [\vec{A}^{\text{ext}}] 
\,,
\end{eqnarray}
so that in the continuum limit $a \rightarrow 0$ and  
$B$ fixed we get the classical energy density 
$B^2/2$.

\section{U(1) FINITE SIZE SCALING}

To further convince ourselves of the consistency of our evaluation 
of the lattice effective action,
we also determined the critical parameters of U(1) l.g.t. by applying
the standard  finite size scaling analysis~\cite{Barber88}.
To this purpose we simulated U(1) l.g.t. on lattices with sizes 
$L_1=64$, $L_2=L_3=L_4\equiv L=6, 8, 10, 12, 14$ and $n^{\text{ext}}=2$.
In
Figure~2 we display the derivative of the energy density 
near the critical region for
different lattice sizes ($n^{\text{ext}}=2$).
The effective linear dimension is given by 
\begin{equation}
\label{Leff}
L_{\text{eff}} = (\Omega_{\text{int}}(L_1,L_2,L_3,L_4))^{1/4}  
\,.
\end{equation}
We apply the f.s.s. to our  generalized susceptibility
$\varepsilon^{\prime} [ \vec{A}^{\text{ext}} ]$. 
Near the
critical region (see Fig.~2):
\begin{equation}
\label{peak}
\varepsilon^{\prime} \left[\vec{A}^{\text{ext}} \right] =
\frac{a_1(L_{\text{eff}})}{a_2(L_{\text{eff}})
[ \beta - \beta^{*}(L_{\text{eff}})]^2 +1}
\end{equation}
According to f.s.s. the peak of the derivative of the U(1) energy density
($a_1(L_{\text{eff}})$ in Eq.~(\ref{peak}) should behave as
\begin{equation}
\label{a1}
a_1(L_{\text{eff}}) = c  L_{\text{eff}}^{\gamma/\nu}  
\,.
\end{equation}
In Fig.~3 the peak of 
$\varepsilon^{\prime} [\vec{A}^{\text{ext}} ]$ vs. lattice size
with superimposed fit Eq.~(\ref{a1}) is displayed.
\begin{figure}[tbp]
\begin{center}
\epsfig{file=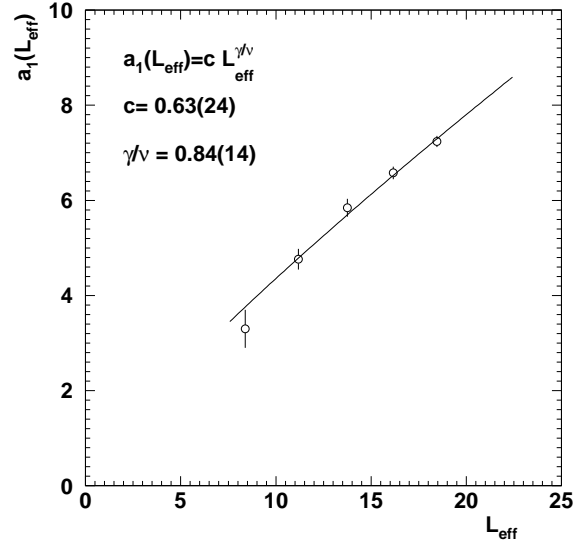,width=7.5truecm}
\vspace{-1.54truecm}
\caption{The maximum of $\varepsilon_{\text{int}}^{\prime}$ obtained
from the fit Eq.~(\ref{peak}) versus $L_{\text{eff}}$ with
superimposed the fit Eq.~(\ref{a1}).}
\end{center}
\end{figure}
Analogously the pseudocritical couplings at various lattice sizes are
fitted according to (see Fig.~4):
\begin{equation}
\label{betac}
\beta^*(L_{\text{eff}}) = \beta_c + k L_{\text{eff}}^{-1/\nu} 
\,.
\end{equation}
\begin{figure}[tbp]
\begin{center}
\epsfig{file=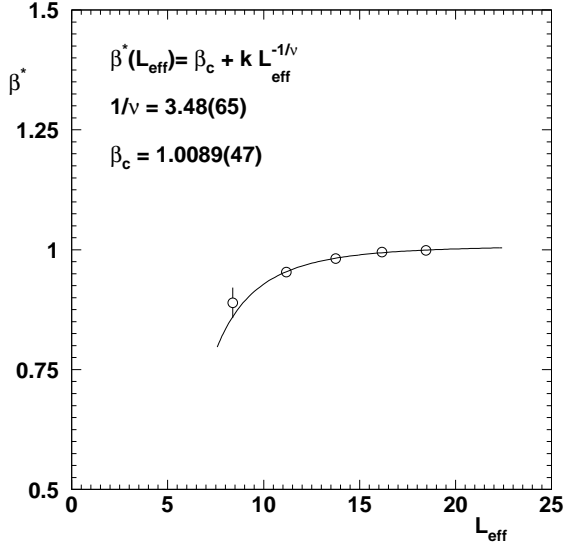,width=7.5truecm}
\vspace{-1.54truecm}
\caption{The pseudocritical couplings at various lattice sizes with
superimposed the fit Eq.~(\ref{betac}).}
\end{center}
\end{figure}
Note that the values of the critical exponents are quite consistent with the
hyperscaling relation:
\begin{equation}
\label{hyperscaling}
\gamma = 2 - \nu d  \Longrightarrow 
\frac{1}{\nu} = \frac{d}{2}+ \frac{1}{2}\frac{\gamma}{\nu} \,.
\end{equation}
Indeed, using the value of $\gamma/\nu=0.84(14)$ obtained from the fit
to the value of the peak of 
$\varepsilon^{\prime}[\vec{A}^{\text{ext}}]$ and the hyperscaling  
relation we get $1/\nu = 2.42 (7)$, which is compatible with
$1/\nu=3.48(65)$ obtained from the fit Eq.~(\ref{betac}) to the
pseudocritical couplings.

We also checked that the 
universality law in the critical region
\begin{equation}
\label{universality}
L_{\text{eff}}^{-\gamma/\nu} \varepsilon_{\text{int}}^{\prime}  \left[
\vec{A}^{\text{ext}} \right] = \tilde{\phi} \left[
L_{\text{eff}}^{1/\nu}(\beta-\beta_c) \right]
\end{equation}
holds quite well for the lattices with $L=10$, $12$, $14$
corresponding to $L_{\text{eff}}=13.75$, $16.16$, $18.46$
respectively (see Fig.~5).
\begin{figure}[tbp]
\begin{center}
\epsfig{file=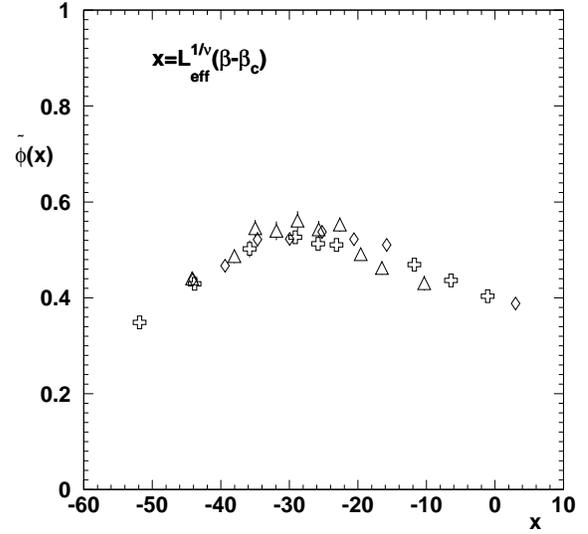,width=7.5truecm}
\vspace{-1.54truecm}
\caption{The universality law Eq.~(\ref{universality}) in the critical
region. Notations as in Fig.~2.
}
\end{center}
\end{figure}

\section{CONCLUSIONS}

We have presented a method that allows to investigate the effective
action for external background fields in gauge systems by means of
Monte Carlo simulations. We have successfully tested this method for
the U(1) pure lattice gauge theory in an external magnetic field. In
particular we  found that the external magnetic field is screened for
strong couplings while penetrates for color weak couplings. We have
also verified that in the continuum limit the effective action
agrees with the classical U(1) action. Moreover our extimations of the
critical parameters and of the infinite volume critical coupling are
in perfect agreement with the values extracted from the specific heat
on lattices with closed topology or with fixed boundary conditions,
where there is evidence of a continuous phase 
transition~\cite{Mutter82,Baig94,Jersak95}.

We used this definition of lattice effective action to analyze the SU(2)
lattice gauge theory in presence of an external magnetic abelian field,
finding evidence for the so-called Nielsen-Olesen unstable
modes~\cite{Lat96}.

\end{document}